\documentclass[showpacs,showkeys,aps,
pra,
 amsmath,amssymb,
preprint,superscriptaddress,%
]{revtex4-1}
\usepackage{graphicx}
\usepackage{dcolumn}
\usepackage{bm}

\newcommand{\beq}{\begin{equation}}
\newcommand{\eeq}{\end{equation}}
\newcommand{\bea}{\begin{eqnarray}}
\newcommand{\eea}{\end{eqnarray}}

\begin{document}

\preprint{AIP/123-QED}

\title{Description of two-electron atoms with correct cusp conditions}

\author{A.T. Kruppa}
\email{kruppa.andras@atomki.mta.hu}
\affiliation{ 
Hungarian Academy of Sciences Institute for Nuclear Physics}
 \author{J. Kov\'acs}
  \email{kovacs.jozsef@atomki.mta.hu}
 \affiliation{ 
Hungarian Academy of Sciences Institute for Nuclear Physics}
\author{I. Hornyak}
\email{ihornyak@atomki.mta.hu}
\affiliation{ Department of Physics, School of Science and Technology,
Nazarbayev University, Astana 010000, Kazakhstan}
\affiliation{ 
Hungarian Academy of Sciences Institute for Nuclear Physics}

\date{\today}

\begin{abstract}
New sets of functions with arbitrary large finite cardinality are constructed 
for two-electron atoms. 
Functions from these sets exactly satisfy the Kato's cusp conditions. 
The new functions are special linear combinations of Hylleraas- and/or Kinoshita-type terms.
Standard variational calculation, leading to matrix eigenvalue problem,  can be carried out 
to calculate the energies of the system. 
There is no need for optimization with constraints to satisfy the cusp conditions. 
In the numerical examples  the ground state energy of the He atom is considered.

\end{abstract}
\maketitle

\section{Introduction}

For a Coulombic system an exact eigenfunction has strange local behaviors namely it has cusps.
The first derivative of the wave function is discontinuous at those points in the configuration space where
two or more charged particles come together. This phenomena is characterized by the Kato's cusp conditions \cite{Kat57}. 
These conditions were also derived for the He atom in \cite{Roo60} and the general treatment was developed in \cite{Pac66}.
A deficiency of the configuration interaction method is that it can not describe the electron-electron (e-e) cusp condition \cite{Roo60,Tew06}. It is not possible to describe the cusp using products of smooth orbital
functions. Under special circumstances explicitly  correlated trial wave functions
can exactly satisfy the cusp conditions.
The importance of the Kato's cusp conditions have been demonstrated several times \cite{Abe70,Dal92,Chu06}. 
For example in the derivation of the double photo-ionization cross section it was assumed that the cusp conditions are fulfilled \cite{Abe70}. 
The role of the cusp conditions is investigated in electron-atom double ionization \cite{Jon03,Anc04,Anc08}.

For the description of $S$ states of  two-electron atoms the standard Hylleraas-variables $s$, $t$ and $u$ are used.
The Hylleraas-type trial function \cite{Hyl29} is a power series expansion in terms of the 
variables $s$, $t$ and $u$. A more general expansion was introduced 
by Kinoshita \cite{Kin57} where negative powers of the  $s$ and $u$ variables 
can appear. The space part of a Kinoshita-type trial wave function  is a finite superposition of basis functions of the form
\begin{equation}\label{kinterm}
w_{l,m,n}(s,t,u)=\exp(-\alpha s)s^l\left(\frac{u}{s}\right)^m
\left(\frac{t}{u}\right)^n.
\end{equation}
Here $l$, $m$ and $n$ are non-negative integers and $\alpha$ is a positive real number.
In this paper, starting from Hylleraas- and Kinoshita-type basis  new sets of functions are constructed in analytic form in order to exactly fulfill the cusp conditions.


There are two types of approaches to get a trial function with correct cusp conditions. Either the mean value of the Hamiltonian-operator is minimalized subject to the cusp conditions \cite{Ten94,Chu06,Hor17} or special basis functions are used in the calculations 
\cite{Rod05,Rod07,Kle96,Sec97,Sch62,Hir63,Bha96,Pat04,Sie99,Plu50,Anc07,Anc07b,Gas08}. 
Very recently a simple but nontrivial Hylleraas-type function is suggested \cite{Car18} which exactly fulfills the cusp conditions. 

In our earlier paper \cite{Hor17} finite terms trial 
wave functions of Hylleraas- or Kinoshita-type were considered and the consequences of the cusp equations are studied. Based on the results of \cite{Hor17} in the present paper first it is shown that the trial wave function of \cite{Car18} can be  obtained very easily from the  formalism of \cite{Hor17}. Our main result is the construction of  new  highly nontrivial function sets which can be composed from finite number of Kinoshita-terms in such a way that  the new functions exactly fulfill  the Kato's cusp conditions.

In section \ref{sect1} the results of the paper \cite{Hor17} relevant to the present work are summarized and methods to fulfill the cusp conditions using interparticle coordinates  
are shortly reviewed. The determination of the new basis functions and a few explicit  examples are contained in section \ref{sect2}. Numerical results are presented in part \ref{sect3} by calculating the ground state energy of the He atom. Finally a summary is given in part \ref{sect4}.

\section{Trial wave functions and cusp conditions }\label{sect1}

The coordinates  of the electrons are denoted by ${\bf r_1}$ and ${\bf r_2}$ and a nucleus with infinite mass  with charge number $Z$ is assumed. 
In the description of the $S$ states of two-electron atoms it is enough to use three scalar variables. 
The Hylleraas-variables are $s=r_1+r_2=\vert{\bf r_1}\vert+\vert{\bf r_2}\vert$, $t=r_1-r_2=\vert{\bf r_1}\vert-\vert{\bf r_2}\vert$ and  $u=r_{12}=\vert{\bf r_{12}}\vert=\vert{\bf r_1}-{\bf r_2}\vert$. The variables $r_1$, $r_2$ and $r_{12}$ are called  interparticle coordinates. In the study of the Kato's cusp conditions mainly interparticle coordinates are used. In the next section 
approaches, where the cusp conditions are exactly fulfilled using special forms for the trial function, are shortly surveyed.

\subsection{Trial functions using interparticle coordinates}

Using finite number of Slater-determinants as in the standard configuration interaction method the 
e-e cusp condition can not be satisfied \cite{Roo60,Tew06}. Explicitly correlated trial functions have to be used. It was suggested in \cite{Roo60} 
the following form for the trial wave function
$
\Phi(r_1,r_2)\chi(r_{12})
$. The closed shell wave function is of the form 
$
\Phi(r_1,r_2)=\phi(r_1)\phi(r_2), 
$
the open shell function looks like
$
\Phi(r_1,r_2)=\phi(r_1)\psi(r_2)+\psi(r_1)\phi(r_2).
$
This type of  wave function  satisfies the cusp equations  if the individual functions $\phi(r)$, $\psi(r)$ and $\chi(r)$ fulfill the conditions  \cite{Roo60}
$\phi'(0)=-Z\phi(0),\ \ \ \psi'(0)=-Z\psi(0)$ and $\chi'(0)=\frac{1}{2}\chi(0)$.
Usually for $\phi(r)$ and $\psi(r)$ the hydrogenic Coulomb-functions are used.

A few suggestions along this line of approach for the correlation function $\chi(r_{12})$ are mentioned in the followings. Abbot and Maslen \cite{Abb86} takes the form
$
\exp(\frac{1}{2}r_{12}).
$
The correlation  function 
$
\left(1+\frac{1}{2}r_{12}\right)
$
can be obtained from the study of the asymptotic form of the exact wave function \cite{Kle96}. For the correlation function, due to the structure of the Hamiltonian, 
it is a natural choice $u_k(r_{12})$, the hydrogenic wave function for an electron in the continuum with energy $k^2/2$ \cite{Plu50}.
This so called Pluvinage-type function can be generalized \cite{Anc07}.
The so called 3C Coulomb wave function is widely used in three body Coulomb scattering calculations. Bound state analog of this type of trial function is recently  introduced \cite{Anc07b,Gas08}. Moreover, by construction, the 3C
wave function fulfills Kato’s cusp conditions at all two-body
coalescence points.
From the analysis of the  asymptotic form of the exact wave function such trial functions were suggested where the use of the hydrogenic Coulomb-functions are avoided \cite{Kle96}.

A general method to construct a wave function with correct Kato's cusp conditions is described in \cite{Rod05,Rod07}. Assume that a function $\Psi^{CF}(r_1,r_2,r_{12})$ with correct cusp conditions is given.
A better trial function can be obtained if the following ansatz is considered
\begin{equation}\label{rod}
\Psi^{CF}(r_1,r_2,r_{12})\times \sum_{{\substack{i,j,k\\i\neq 1, j\neq 1, k\neq 1}}}C_{i,j,k} r_1^i r_2^j r_{12}^k,
\end{equation}
where $i,j,k$ are non-negative integers. If the above restrictions are made in the summation then (\ref{rod}) also satisfies the cusp conditions. In the works \cite{Rod05,Rod07} for the function $\Psi^{CF}(r_1,r_2,r_{12})$ the expression 
$e^{-Z (r_1+r_2)}\left(1-\frac{1}{1+2\lambda}e^{-\lambda r_{12}}\right )$ was used. The latter form of its own was used as a trial function in  \cite{Bha96}.
Interesting expressions for the trial wave function were suggested in \cite{Sec97} and  \cite{Sie99}. They are of the form (\ref{rod}) for  
the $\Psi^{CF}(r_1,r_2,r_{12})$ the function 
$e^{-Z (r_1+r_2)}
\left(1+\frac{1}{2}r_{12}e^{-\lambda r_{12}}\right)$ is taken which was suggested in \cite{Hir63}.

\subsection{Hylleraas- and Kinoshita-type trial functions}

For space part of the wave function of a two-electron atom the following form is taken
\begin{equation}\label{kwf}
\phi(s,t,u)=\sum_{l,m,n}k_{l,m,n}w_{l,m,n}(s,t,u). 
\end{equation}
The terms in this trial function are characterized by a triplet of  non-negative integers. The notation 
$[l,m,n]$ is used for such a triplet.  
When an $[l,m,n]$ term is mentioned it means the function $w_{l,m,n}(s,t,u)$. For wave functions with singlet spin part $n$ is even.
If the restriction $l\ge m\ge n$ is used  such a wave function is gained which was suggested by Hylleraas \cite{Hyl29}.
Since its introduction the Hylleraas-type form of variational trial functions have huge number of successful applications. The general form  (\ref{kwf}) is due to Kinoshita \cite{Kin57}. The characteristic of this form is that negative powers of the variables $s$ and $u$ are allowed.

In the rest of this section  those results of \cite{Hor17} are collected which are used later in  the  paper.
The singularities for the electron-nucleus (e-n) coalescences are at the points $(s,-s,s)$ and $(s,s,s)$.
The e-e coalescences occur at the points $(s,0,0)$. The following notation is used: a triplet of numbers in parentheses  corresponds 
to the $s$, $t$ and $u$ values. The Kato's cusp conditions in Hylleraas-coordinates 
are given by the equations \cite{Hor17,Car18}
\begin{equation}\label{K1}
\phi_{s}(s,-s,s)+\phi_{t}(s,-s,s)=-Z\phi(s,-s,s),
\end{equation}
\begin{equation}\label{K2}
\phi_{s}(s,s,s)-\phi_{t}(s,s,s)=-Z\phi(s,s,s)
\end{equation}
and  
\begin{equation}\label{K12}
\phi_{{u}}(s,0,0)=\frac{1}{2}\phi(s,0,0).
\end{equation}
Here the standard mathematical notation is used for the partial derivatives with respect to $s$, $t$ and $u$.

The coefficients $k_{l,m,n}$ of the trial  wave function (\ref{kwf})  have to obey  certain equations in order to satisfy the cusp conditions. The e-n cusp condition can be expressed by 
\begin{equation}\label{katoa1}
\sum_{m,n}(m+n)k_{0,m,n}=0
\end{equation}
and
\begin{equation}\label{katoa2}
\sum_{m,n}\left[(m+n-l)k_{l,m,n}+\bar\alpha k_{l-1,m,n}\right]=0,\ \ l>0.
\end{equation}
Here the abbreviation $\bar\alpha =\alpha -Z $ is introduced.
The fulfillment of the e-e cusp condition leads to 
$k_{0,1,0}=0$
and
\begin{equation}\label{katob2}
k_{l,1,0}=\frac{1}{2}k_{l-1,0,0},\ \ l>0.
\end{equation}
For finite terms wave functions two more groups of constraints present. The restrictions
$k_{l,0,n}=0,\ \  n>0$
assure the limit of the trial wave function at the singularity points $(s,0,0),\ \ s\neq 0$. The second group of restrictions
$k_{l,1,n}=0,\ \  n> 1$
ensures the limit of $\phi_u$ at
the e-e coalescence line $(s,0,0),\ \ s>0$.
To satisfy the cusp conditions in the triple coalescence point $(0,0,0)$  the terms with $l=0$ and $l=1$
are severely restricted. The only possible terms are $[0,0,0], [1,1,0]$ and $[1,0,0]$.

\section{Basis functions with exact cusp conditions}\label{sect2}

Using the $l=0$ restriction i.e. for $l=0$ the only possible term is $[0,0,0]$, Eq. (\ref{katoa1}) turns into 
$0\ k_{0,0,0}=0$ which 
can be fulfilled for arbitrary $k_{0,0,0}$. 

First it is assumed that $\bar\alpha=0$, in this case the e-n cusp conditions are simplified. From (\ref{katoa2}) it follows that
\begin{equation}\label{katonewa2}
\sum_{m,n}(m+n-l)k_{l,m,n}=0,\ \ l\geq 1.
\end{equation}
Using the $l=1$ restriction i.e. for $l=1$ the only possible terms are $[1,0,0]$ and $[1,1,0]$, from Eqs. (\ref{katonewa2}) and (\ref{katob2}) 
it can be deduced
$ k_{1,0,0}=k_{2,1,0}=0$.

\subsection{Simple solutions}\label{simpsol}
A trivial solution of the e-n cusp equations (\ref{katonewa2}) can be obtained if such terms are used only where $l=m+n$ and $\bar\alpha=0$. Because of these strict restrictions only the $l=1$ e-e cusp condition has to be considered and  according to  (\ref{katob2})  $k_{1,1,0}=k_{0,0,0}/2$. In such a circumstances the trial function satisfying all cusp conditions  can be written into the form 
\begin{equation}\label{simple1} 
 k_{0,0,0}\left(w_{0,0,0}+\frac{1}{2}w_{1,1,0}\right)+\sum_{{\substack{l,m,n\\l=m+n, l>1}}} k_{l,m,n}w_{l,m,n}.
\end{equation}
To save space the $s$, $t$ and $u$ arguments of the functions $w_{l,m,n}(s,t,u)$ are not shown.
The restriction $l=m+n$ means that the form of the basis functions are $w_{m+n,m,n}=\exp(-Z s)s^nu^{m-n}t^n$. 
In the case of Kinoshita-type function this means that only $u$ may have negative exponent and the powers of  $s$ and $t$ are the same.
If Hylleraas-type function is considered it can be written that $m=n+k$ ($0\leq k$) so the allowed
function form is $w_{2n+k,n+k,n}=\exp(-Z s)s^nu^kt^n$. Separating the $n=0$ term from the rest (\ref{simple1}) can be rewritten
\begin{equation}\label{simple2} 
\exp(-Z s)\left[ k_{0,0,0}\left(1+\frac{1}{2}u\right)+
\sum_{\substack{n,k\\n>0}} k_{2n+k,n+k,n}s^n u^k t^n+\sum_{\substack{ k\\k>1}}k_{k,k,0} u^k \right].
\end{equation}
The third term in (\ref{simple2}) can be neglected and still the cusp conditions are fulfilled since it is not necessary to use all  functions with the condition $l=m+n$. In this case the trial function is 
\begin{equation}\label{simple3} 
\exp(-Z s)\left[ k_{0,0,0}\left(1+\frac{1}{2}u\right)+\sum_{\substack{n,k\\n>0}} k_{2n+k,n+k,n}s^n u^k t^n\right].
\end{equation}
This last form was derived in a recent paper \cite{Car18} using other reasoning. It can be expected that the very simple forms of  the trial functions (\ref{simple2}) and (\ref{simple3}) would  result in not very accurate energy eigenvalue.
In the rest of this section the restriction $\bar\alpha=0$ is overcome and 
the simplicity of the allowed terms is surmounted.

\subsection{Basis functions with exact cusp conditions}
Let's assume that a trial wave function is given in the form  
\begin{equation}\label{proba1}
\sum_{l=0}^Lk_{l,0,0}(w_{l,0,0}+\frac{1}{2}w_{l+1,1,0})+\sum'_{l,m,n} k_{l,m,n} w_{l,m,n}(s,t,u).
\end{equation}
The sign $'$ above the summation means that $[l,m,n]\ne [l,0,0]$ and  $[l,m,n]\ne [l,1,0]$.  Here the general case is considered i.e. $\bar\alpha\neq 0$. Since a finite term wave function is looked for, the values of $l$ are restricted, and the maximum of the values of $l$   
is denoted by $L$. It is assumed that for a given $l$ only finite number of  $m$ and $n$ values are taken into account.
The special summation notation also means the summation over $l$ in the second summation of (\ref{proba1}) runs between $l=2$ and $l=L$.  

Considering (\ref{katob2}) it is obvious that (\ref{proba1}) satisfies the e-e cusp conditions for $l=1,\ldots ,L$.
The coefficients $k_{l,m,n}$ are arbitrary in the second summation in (\ref{proba1}) whereas the $k_{l,0,0}$ coefficients in the first term 
of (\ref{proba1}) are dependent ones. They can not be determined  variationally if such a wave function is looked for where the cusp conditions are exactly satisfied. 
A recursive solution of the cusp equations is given in \cite{Hor17} the coefficients $k_{l,0,0}$ are given by
\bea\label{cuspsolz}
k_{l+2,0,0}=B_{l+2}+\frac{1}{l+2}\left(\bar\alpha -\frac{l+1}{2}\right)k_{l+1,0,0}
+\frac{\bar\alpha }{2(l+2)}k_{l,0,0} ~~~l\ge 0
\eea
where
\begin{equation}\label{bdef}
B_l=\frac{1}{l}\sum _{m>1,n}\left [\left ( m+n-l\right )k_{l,m,n}+\bar\alpha k_{l-1,m,n}\right ]\ \ l\geq 2.
\end{equation}
The initial conditions for the recursion are: $k_{0,0,0}$ is arbitrary and 
$k_{1,0,0}=\bar\alpha  k_{0,0,0}$.
This last initial condition stems from (\ref{katoa2}) when $l=1$.  
The explicit form of $k_{l,0,0}$ reads
\begin{eqnarray}\label{expsolz}
&k_{l,0,0}=\frac{\bar\alpha ^l}{l!}k_{0,0,0}+\frac{(-1)^l}{l}\frac{1}{2^{l}}
\sum_{i=2}^{l}i (-1)^i 2^{i} B_i\nonumber\\
&+\frac{(-1)^l}{l!2^l}\sum_{k=1}^{l-1}(-1)^k \bar\alpha ^k(l-k-1)!2^{k}
\sum_{i=2}^{l-k}i (-1)^i 2^{i} B_i\ \ \ l\geq 2.
\end{eqnarray}
Here the convention is used that if in a summation the lower bound is larger than the upper one then the value of the summation is zero. The proof of (\ref{expsolz}) is given in the Appendix \ref{proof}. 
The recursive solution (\ref{cuspsolz}) and (\ref{expsolz}) are valid if the e-e cusp conditions (\ref{katob2}) are also fulfilled.

A set of integer triplets $\cal D$ is introduced  it contains 
the dependent  variables  ${\cal D}=\{[l,0,0], [l,1,0]\vert 1\le l \le L\}\cup \{[L+1,1,0]\}$.
The set of the integer triplets $[l,m,n]$ appearing in the second summation of (\ref{proba1}) 
together with  the special integer triplet $[0,0,0]$ is denoted by $\cal F$. 
Earlier at the beginning of section \ref{sect2} it was found that the cusp conditions do not 
fix the value of $k_{0,0,0}$. The set $\cal F$  contains the free, independent $k_{l,m,n}$ expansion coefficients.  

If the solution (\ref{expsolz}) is used in (\ref{proba1}) then all cusp conditions are fulfilled for $0\le l \le L$. 
Unfortunately the term $[L+1,1,0]$ is  
present in (\ref{proba1}) so the the coupled e-n cusp equations (\ref{katoa2}) for $l=L+1$ and $l=L+2$  have to be explicitly considered and solved. 
This can be achieved if extra terms $[L+1,m,n]$ or $[L+2,m,n]$ are added to (\ref{proba1}). The $m$ and $n$ indexes of these extra terms are called auxiliary parameters.
If  trial function with minimal number of terms are requested
then the term $[L+1,0,0]$ could not be added to the trial function (\ref{proba1}). 

There is large freedom how to satisfy the e-n cusp conditions for  
$l=L+1$ and $l=L+2$.  Two simple cases are considered. Two $l=L+1$ terms can be added to  (\ref{proba1}) they are  denoted by 
$[L+1,m_0,n_0]$ and $[L+1,m_1,n_1]$. An alternative way is to add three extra terms to (\ref{proba1}) and they are signed by 
$[L+1,m_0,n_0]$,  $[L+2,m_\alpha,n_\alpha]$ and $[L+2,m_\beta,n_\beta]$. In this case the e-n cusp equation for $l=L+3$ has to be also considered
but with the restriction that  terms with $l\ge L+3$ are not in the trial function. We have to exclude the terms 
$[L+2,1,0]$ and $[L+2,0,0]$ from the selected ones since
we do not want to bother about new e-e cusp conditions and do not want terms such that $l\ge L+3$. 

In the first case, when to extra terms are added to (\ref{proba1}),
the two  coupled e-n cusp equations  (\ref{katoa2})  for $l=L+1$ and $l=L+2$ can be solved for the variables
  $k_{L+1,m_0,n_0}$,  $k_{L+1,m_1,n_1}$ and the results can be put into the form  
\begin{eqnarray}\label{lp1b}
& k_{L+1,m_0,n_0}=\frac{1}{m_0+n_0-m_1-n_1}\left( \left(\frac{m_1+n_1-1}{2}-\bar\alpha \right) k_{L,0,0}
-\frac{\bar\alpha }{2}k_{L-1,0,0}\right)\nonumber\\
&-\frac{\bar\alpha }{m_0+n_0-m_1-n_1}\sum'_{m,n}k_{L,m,n} 
\end{eqnarray}
and
\begin{eqnarray}\label{lp2b}
& k_{L+1,m_1,n_1}=-\frac{1}{m_0+n_0-m_1-n_1}\left( \left(\frac{m_0+n_0-1}{2}-\bar\alpha \right) k_{L,0,0}
-\frac{\bar\alpha }{2}k_{L-1,0,0}\right)\nonumber\\
&+\frac{\bar\alpha }{m_0+n_0-m_1-n_1}\sum'_{m,n}k_{L,m,n}. 
\end{eqnarray}
The solutions are written down such a way that the dependent variables are separated out and the e-e 
cusp conditions are considered.
The set of dependent variables is  
${\cal D}=\{[l,0,0],[l,1,0]\vert 1\le l \le L\}\cup \{[L+1,m_0,n_0],[L+1,m_1,n_1],[L+1,1,0]\}$.
Here it is required that $m_0+n_0-m_1-n_1\ne 0$.  
Although there is no summation over $l$ in (\ref{lp1b}) and in (\ref{lp2b}) the special summation notation has the same meaning as before since it is obvious what is 
the value of $l$.
The final form of the wave function which fulfills the Kato's cusp conditions is
\begin{eqnarray}\label{probafullz}
&\sum'_{l,m,n}k_{l,m,n} w_{l,m,n}
+\sum_{{\substack{l=0}}}^Lk_{l,0,0}\left(w_{l,0,0}+\frac{1}{2}w_{l+1,1,0}\right)\nonumber\\
&+k_{L+1,m_0,n_0}w_{L+1,m_0,n_0}
+k_{L+1,m_1,n_1}w_{L+1,m_1,n_1}.
\end{eqnarray}

In the second case, when three extra terms are added to (\ref{proba1}), 
the three  coupled e-n cusp equations  (\ref{katoa2}) for $l=L+1$, $l=L+2$ and $l=L+3$ can be solved for the variables  $k_{L+1,m_0,n_0}$,  $k_{L+2,m_\alpha,n_\alpha}$ and
$k_{L+2,m_\beta,n_\beta}$. They can be expressed as
\begin{equation}\label{lp1a}
k_{L+1,m_0,n_0}=\frac{1}{m_0+n_0-L-1}\left(\left (\frac{L}{2}-\bar\alpha \right)k_{L,0,0}-\bar\alpha \left(\frac{1}{2}k_{L-1,0,0}+\sum'_{m,n}k_{L,m,n}\right)\right),
\end{equation}
\begin{eqnarray}\label{lp2a}
&k_{L+2,m_\alpha,n_\alpha}=-\frac{\bar\alpha }{(m_0+n_0-L-1) (m_\alpha+n_\alpha-m_\beta-n_\beta)} \nonumber\\
&\times\left(\left(\frac{1}{2}(m_0+n_0-1)-\bar\alpha \right)k_{L,0,0}-\bar\alpha \left(\frac{1}{2}k_{L-1,0,0}+\sum'_{m,n}k_{L,m,n}\right)\right)
\end{eqnarray}
and 
$
k_{L+2,m_\beta,n_\beta}=-k_{L+2,m_\alpha,n_\alpha}.
$
For the selection of the extra terms the restrictions are $m_0+n_0-L-1\ne 0$ and $m_\alpha+n_\alpha-m_\beta-n_\beta\ne 0$. The set of the  dependent variables is ${\cal D}=\{[l,0,0],[l,1,0]\vert 1\le l \le L\}\cup \{[L+1,m_0,n_0],[L+2,m_\alpha,n_\alpha],[L+2,m_\beta,n_\beta],[L+1,1,0]\}$.
In the second case the final form of the wave function which fulfills the Kato's cusp conditions is
\begin{eqnarray}\label{probafullz2}
&\sum'_{l,m,n}k_{l,m,n} w_{l,m,n}
+\sum_{{\substack{l=0}}}^Lk_{l,0,0}\left(w_{l,0,0}+\frac{1}{2}w_{l+1,1,0}\right)\nonumber\\
&+k_{L+1,m_0,n_0}w_{L+1,m_0,n_0}
+k_{L+2,m_\alpha,n_\alpha}w_{L+2,m_\alpha,n_\alpha}+k_{L+2,m_\beta,n_\beta}w_{L+2,m_\beta,n_\beta}.
\end{eqnarray}

The most important point of our method is to rewrite the trial functions  (\ref{probafullz}) and (\ref{probafullz2}) which  exactly fulfill the cusp conditions into the following form 
\begin{equation}\label{final1}
\sum_{[l,m,n]\in {\cal F}}k_{l,m,n}\tilde w_{l,m,n}(s,t,u).
\end{equation}
To do this the set of the dependent variables $\cal D$  have to be taken into account .
Substituting the values of the dependent variables into (\ref{probafullz}) and  (\ref{probafullz2})  the coefficients 
$\tilde w_{l,m,n}(s,t,u)$ of the free expansion variables $k_{l,m,n}$ can be collected. The coefficients of the free $k_{l,m,n}$ variables are functions of the Hylleraas-coordinates. 
These coefficients  define the new basis functions $\tilde w_{l,m,n}(s,t,u)$. 
To do this task easily it is good to know what is the coefficient of the free variable $k_{l,m,n}$ in $k_{r,0,0}$. This is denoted by $C_r(l,m,n)$ and it is given for $r\geq 2$ by the expression

\begin{eqnarray}\label{cexp}
&C_r(l,m,n)=(m+n-l)\frac{(-1)^{l+r}2^{l-r}}{r!}\left((r-1)!+\sum_{k=1}^{r-l}(-1)^k\bar\alpha ^k(r-k-1)!2^k\right)\nonumber\\
&-\bar\alpha \frac{(-1)^{l+r}2^{l-r+1}}{r!}\left((r-1)!+\sum_{k=1}^{r-l-1}(-1)^k\bar\alpha ^k(r-k-1)!2^k\right)
\end{eqnarray}
if $[l,m,n]\in {\cal F}\setminus \{[0,0,0]\}$ and $r\geq l+1$.  If $r=l$ only the first term should be used 
for the calculation of $C_r(l,m,n).$ Of course if $[l,m,n]\not\in {\cal F}$ or $r<l$ then $C_r(l,m,n)=0$. It is easy to see from (\ref{expsolz}) that
\begin{equation}
C_r(0,0,0)=\frac{\bar\alpha ^r}{r!}\ \ \ r\geq 2.
\end{equation}

If the trial function (\ref{probafullz}) is considered the coefficient of $k_{0,0,0}$ is
\begin{eqnarray}\label{000za}
&\tilde w_{0,0,0}(s,t,u)=w_{0,0,0}+\frac{1}{2}w_{1,1,0}+\sum_{l=1}^L\frac{\bar\alpha ^l}{l!}\left(w_{l,0,0}+\frac{1}{2}w_{l+1,1,0}\right)\nonumber\\
&+\frac{w_{L+1,m_0,n_0}\bar\alpha ^{L}}{(m_0+n_0-m_1-n_1)L!}\left(\frac{1}{2}(m_1+n_1-1)-\bar\alpha -\frac{1}{2}L\right)\nonumber\\
&-\frac{w_{L+1,m_1,n_1}\bar\alpha ^{L}}{(m_0+n_0-m_1-n_1)L!}\left(\frac{1}{2}(m_0+n_0-1)-\bar\alpha -\frac{1}{2}L\right).
\end{eqnarray}
The coefficients of the other free variables  $k_{l,m,n}$ read
\begin{eqnarray}\label{lmnzterma}
&\tilde w_{l,m,n}(s,t,u)=w_{l,m,n}+\sum_{i=l}^L C_i(l,m,n)\left(w_{i,0,0}+\frac{1}{2}w_{i+1,1,0}\right)\nonumber\\
&+\frac{w_{L+1,m_0,n_0}}{m_0+n_0-m_1-n_1}\left((\frac{1}{2}(m_1+n_1-1)-\bar\alpha )C_L(l,m,n)-\bar\alpha (\frac{1}{2}C_{L-1}(l,m,n)+\delta_{l,L})\right)\nonumber\\
&-\frac{w_{L+1,m_1,n_1}}{m_0+n_0-m_1-n_1}\left((\frac{1}{2}(m_0+n_0-1)-\bar\alpha )C_L(l,m,n)-\bar\alpha (\frac{1}{2}C_{L-1}(l,m,n)+\delta_{l,L})\right).
\end{eqnarray}

If the trial function (\ref{probafullz2}) is considered the coefficient of  $k_{0,0,0}$ is 
\begin{eqnarray}\label{000zb}
&\tilde w_{0,0,0}(s,t,u)=w_{0,0,0}+\frac{1}{2}w_{1,1,0}+\sum_{l=1}^L\frac{\bar\alpha ^l}{l!}\left(w_{l,0,0}+\frac{1}{2} w_{l+1,1,0}\right)-w_{L+1,m_0,n_0}\frac{\bar\alpha ^{L+1}}{(m_0+n_0-L-1)L!}\nonumber\\
&-\frac{\bar\alpha ^{L+1}(w_{L+2,m_\alpha,n_\alpha}-w_{L+2,m_\beta,n_\beta})}{(m_0+n_0-L-1)
(m_\alpha+n_\alpha-m_\beta-n_\beta)L!}\left(\frac{1}{2}(m_0+n_0-L-1)-\bar\alpha \right)
\end{eqnarray}
and the coefficients of the other independent variables are
\begin{eqnarray}\label{lmnztermb}
&\tilde w_{l,m,n}(s,t,u)=w_{l,m,n}+\sum_{i=l}^L C_i(l,m,n)\left(w_{i,0,0}+\frac{1}{2}w_{i+1,1,0}\right)\nonumber\\
&+\frac{w_{L+1,m_0,n_0}}{m_0+n_0-L-1}\left((\frac{L}{2}-\bar\alpha )C_L(l,m,n)-\bar\alpha (\frac{1}{2}C_{L-1}(l,m,n)+\delta_{l,L})\right)\nonumber\\
&-\frac{(w_{L+2,m_\alpha,n_\alpha}-w_{L+2,m_\beta,n_\beta})\bar\alpha }{(m_0+n_0-L-1)(m_\alpha+n_\alpha-m_\beta-n_\beta)}\nonumber\\
&\left((\frac{1}{2}(m_0+n_0-1)-\bar\alpha )C_L(l,m,n)-\bar\alpha (\frac{1}{2}C_{L-1}(l,m,n)+\delta_{l,L})\right).
\end{eqnarray}

The function $\tilde w_{l,m,n}(s,t,u)$ is called cusp function (CF). If the forms (\ref{000za}) - (\ref{lmnzterma}) and  (\ref{000zb}) - (\ref{lmnztermb}) 
are used than it is called first- and second-type CF. In both cases the number of terms in the final trial function (\ref{final1}) is the same. 
The number of CF's is the cardinality of the set $\cal F$. 
It is obvious from our derivation that the set of the cusp functions are determined by $L$, the auxiliary parameters and the set $\cal F$. 
Any change in these parameters modifies the set of the CF's.  Let's assume that $L$ and the auxiliary parameters are fixed. It follows from (\ref{000za}) and (\ref{000zb})
that $\tilde w_{0,0,0}(s,t,u)$ is independent of $\cal F$. Assume that there are two sets of free parameters $\cal F$ and ${\cal F}'$. The 
expressions (\ref{cexp}), (\ref{lmnzterma}) and (\ref{lmnztermb})  show that if $[l,m,n]\in {\cal F}\cap {\cal F}'$ then in both CF sets the functions $\tilde w_{l,m,n}(s,t,u)$ belong to $\cal F$ and $\cal F'$ agree with each others.

According to our derivation the function $(\ref{final1})$ exactly fulfills the cusp conditions. 
It can be shown that this is true also for any individual CF $\tilde w_{l,m,n}(s,t,u)$. First  $\tilde w_{0,0,0}(s,t,u)$ is considered. Let's take ${\cal F}=\{[0,0,0]\}$ than  $\tilde w_{0,0,0}(s,t,u)$ agrees with (\ref{final1}) and this proves that  $\tilde w_{0,0,0}(s,t,u)$ satisfies the cusp conditions. For a given $L$ and given auxiliary 
parameters the CF $\tilde w_{0,0,0}(s,t,u)$ does not depend on the set $\cal F$. Now the set ${\cal F}=\{[0,0,0],[l,m,n]\}$ is considered.
In this case (\ref{final1}) turns into  $k_{0,0,0}\tilde w_{0,0,0}(s,t,u)+k_{l,m,n}\tilde w_{l,m,n}(s,t,u)$, since this superposition and  $\tilde w_{0,0,0}(s,t,u)$ fulfill the cusp conditions  it is easy to see that $\tilde w_{l,m,n}(s,t,u)$ also satisfies  the differential cusp conditions (\ref{K1}), (\ref{K2}) and (\ref{K12}).

In the special case $\bar\alpha=0$ the expressions of the CF's are much simpler than the general case. The function (\ref{cexp}) for $r\geq 2$ has a  simple form  $C_r(l,m,n)=(m+n-l)(-1)^{l+r}2^{l-r}/r$ if $[l,m,n]\in {\cal F}\setminus \{[0,0,0]\}$ and $r\geq l$ otherwise its value is zero and obviously $C_r(0,0,0)=0$. 
Both for the  first- and second-type CF  the function $\tilde w_{0,0,0}(s,t,u)$ has the same form and  according to  (\ref{000zb}) and (\ref{000za}) it is given by 
\begin{equation}
\tilde w_{0,0,0}(s,t,u)=w_{0,0,0}+\frac{1}{2}w_{1,1,0}=\exp(-Zs)\left( 1+\frac{1}{2}u\right).
\end{equation}
The other first type CF looks like
\begin{eqnarray}\label{neww1}
&\tilde w_{l,m,n}(s,t,u)=w_{l,m,n}+(m+n-l)(-1)^l 2^{l}\sum_{i=l}^{L}\frac{(-1)^i}{i 2^{i}}\left( w_{i,0,0}+\frac{1}{2}w_{i+1,1,0}\right)\nonumber\\
&+(m+n-l)(-1)^{l+L}2^{l-L}\frac{w_{L+1,m_0,n_0}}{m_0+n_0-m_1-n_1}\frac{1}{2L}(m_1+n_1-1)\nonumber\\
&-(m+n-l)(-1)^{l+L}2^{l-L}\frac{w_{L+1,m_1,n_1}}{m_0+n_0-m_1-n_1}\frac{1}{2L}(m_0+n_0-1).
\end{eqnarray}
The second-type CF reads
\begin{eqnarray}\label{neww2}
&\tilde w_{l,m,n}(s,t,u)=w_{l,m,n}+(m+n-l)(-1)^l 2^{l}\sum_{i=l}^{L}\frac{(-1)^i}{i 2^{i}}\left( w_{i,0,0}+\frac{1}{2}w_{i+1,1,0}\right)\nonumber\\
&+(m+n-l)
\frac{(-1)^{l+L}\ 2^{l-L-1}}{m_0+n_0-L-1} w_{L+1,m_0,n_0}.
\end{eqnarray}
Notice that if $l=m+n$ and $\bar\alpha=0$ then the CF $\tilde w_{m+n,m,n}(s,t,u)$ agrees with the original Kinoshita-term 
$\tilde w_{m+n,m,n}(s,t,u)=w_{m+n,m,n}$.

Our approach has two very important aspects. First the number of CF's can be arbitrary large, the number of terms in the trial function can be arbitrary  increased in order to get better energy. The CF set can be changed versatilely by modifying the set $\cal F$.
Second, the search for the optimum values for the free $k_{l,m,n}$ variables in (\ref{final1}) leads to the usual matrix eigenvalue problem 
but the matrix elements have to be calculated with the CF's. There is no need to carry out optimization with constraints to satisfy the cusp conditions.

\subsection{Cusp functions for $L=1$ and $L=2$}

The explicit analytic forms of the CF's for the simplest cases are given in this section.
The exponential part of the CF can be separated by the following definition
\begin{equation}
\tilde w_{l,m,n}(s,t,u)=\exp(-\alpha s)\tilde P_{l,m,n}(s,t,u).
\end{equation}\label{dset}
In the rest of this section only the function  $\tilde P_{l,m,n}(s,t,u)$ is written down.
Here it is assumed that only Hylleraas-type terms are used in the construction of CF's. 
The following values are used for the auxiliary parameters: $m_0=2, n_0=2$, $m_1=m_\alpha=2$, $n_1=n_\alpha=0$ and $m_\beta=3, n_\beta=0$.

First  the simplest case $L=1$ is considered, in this case there is only one choice for the set $\cal F$ namely ${\cal F}=\{[0,0,0]\}$.
The first-type CF is
\begin{equation}\label{k000}
\tilde P_{0,0,0}(s,t,u)=\frac{1}{2} \left(s u \bar\alpha +2 s \bar\alpha -t^2 \bar\alpha ^2+u^2 (\bar\alpha -1) \bar\alpha +u+2\right).
\end{equation}
The choices  $m_0=2, n_0=2$ and $m_1=2, n_1=0$ are mandatory. In this case $L+1=2$ and there are four possible integer pairs 
from which $(m_0,n_0)$ and $(m_1,n_1)$ can be selected as 
auxiliary parameters but only the ones mentioned above fulfills the restrictions. 
The second-type cusp function looks like for $L=1$
\begin{equation}
\tilde P_{0,0,0}(s,t,u)=\frac{1}{2} \left(-\bar\alpha ^2 \left(u^2 (u-s)+t^2\right)+u^2 \bar\alpha ^3 (u-s)+s (u+2) \bar\alpha +u+2\right).
\end{equation}
The choice $m_0=2, n_0=2$ is obligatory because of the same restrictions as before and in addition the expression  $m_0+n_0-2\ne 0$ have to be taken into account
so the choice  $m_0=2, n_0=0$ is not allowed.

Next  the $L=2$ case is considered. Here the explicit forms of the CF's are given only for the first-type CF's. In the case $L=2$ there are four different sets of CF's depending on the set $\cal F$. 
In the simplest case ${\cal F}=\{[0,0,0]\}$ and there is only one CF
\begin{eqnarray}\label{l2000}
&\tilde P_{0,0,0}(s,t,u)=\frac{1}{8} \left(s \bar\alpha  \left(4 s \bar\alpha +t^2 (-(2 \bar\alpha +1)) \bar\alpha +8\right)+s u^2 (2 \bar\alpha -1) \bar\alpha ^2\right.\nonumber\\
&\left. +2 u (s \bar\alpha  (s \bar\alpha +2)+2)+8\right).
\end{eqnarray}
Notice that this function is different from (\ref{k000}). 

If the set of the free variables is ${\cal F}=\{[0,0,0],[2,2,0]\}$ then two CF's exist. The function $\tilde P_{0,0,0}(s,t,u)$ is given by (\ref{l2000}) and
\begin{equation}\label{l2220}
\tilde P_{2,2,0}(s,t,u)=\frac{1}{2} s \bar\alpha  \left(u^2-t^2\right)+u^2.
\end{equation}

If the set  ${\cal F}=\{[0,0,0],[2,2,2]\}$ is selected for the independent variables  two cusp functions can be constructed. The function $\tilde P_{0,0,0}(s,t,u)$ is given by (\ref{l2000}) and
\begin{equation}\label{l2222}
\tilde P_{2,2,2}(s,t,u)=\frac{1}{4} s \left(t^2 (1-4 \bar\alpha )+u^2 (4 \bar\alpha -3)+2 s (u+2)\right)+t^2.
\end{equation}

Finally if all possible allowed terms are used  ${\cal F}=\{[0,0,0],[2,2,0],[2,2,2]\}$ then there are three CF's. They are given by (\ref{l2000}), (\ref{l2220}) and 
(\ref{l2222}).

\section{Numerical results}\label{sect3}
\begin{center}
\begin{table}\label{tenergy}
\caption{The ground state energy ($E$) of the He atom in atomic units using different trial wave functions. For detailed explanations see the text. The set of the CF's is characterized by the parameter $L$ and the best auxiliary parameters are shown. The number of linear variational parameters for the models are $N$. The exact ground state energy is -2.903724377 a.u. \cite{Dra02}. }
\begin{tabular}{|l|c|l||c|l||c|c|c|c|c|l||c|c|c|l|}
\hline
L&\multicolumn{2}{|c|}{CF-S1} &\multicolumn{2}{c|}{CF-S2}  &\multicolumn{6}{c|}{CF-Z-I}&\multicolumn{4}{c|}{CF-Z-II} \\
\hline
& E         & N         & E         & N       &$m_0$&$n_0$&$m_1$&$n_1$& E          & N&$m_0$&$n_0$& E&N \\
\hline
2 & -2.876582 &1   & -2.878545 & 2 &2&0&2& 2&-2.898783&3  &2 &2& -2.902970 & 3   \\
3 & -2.876582 &1   & -2.879315 & 3 &2&0&3&0  &-2.903484&7 &2 &0&-2.903432 & 7   \\
4 & -2.887268 &2   & -2.887941 & 5 &2&0&3&0  &-2.903641& 14  &4 &2&-2.903653 & 14  \\
5 & -2.888445 &3   & -2.890252 & 7 &2&0&3&0  &-2.903703& 24&5 &2&-2.903700 & 24  \\
6 & -2.889158 &4    & -2.891359 & 9 &2&0&3&0  &-2.903716&38 &6&2&-2.903715 & 38 \\
7 & -2.889509 &5    &-2.891822  & 11 &2&0&3&0 &-2.903721 &56&7 &2& -2.903720& 56  \\
8&-2.889803&7&-2.892027&14&2&0&3&0&-2.903722&79&8&2&-2.903722&79\\
\hline
\end{tabular}
\end{table}
\end{center}

In this section the ground state energy of the He atom is calculated  using the CF's. Such a CF's are considered where only Hylleraas-type elementary terms are used for the superpositions.
For completeness, results when the simple solutions (\ref{simple2}) and (\ref{simple3}) are used as trial functions, are also given.
A CF calculation can be characterized by the value of $L$. In the so called simple cases one has to keep in mind that in (\ref{simple2}) and (\ref{simple3}) the summations are over $k$ and $n$ and they are restricted by the condition $L=2n+k$.

In a numerical calculation if the trial function is given in the forms (\ref{simple2}) and (\ref{simple3}) they are  called CF-S2 and CF-S1 models, respectively. If the first- and the second-type CF's are used they are referred as CF-$\alpha$-I and CF-$\alpha$-II calculations, respectively. In the case of the models CF-$\alpha$-I and CF-$\alpha$-II the value of the parameter 
$\alpha$ is optimized and also the best values of the auxiliary parameters are  determined. 
If the $\alpha=Z$ choice is made for the first- and second-type CF's then the calculations are called CF-Z-I and CF-Z-II, respectively. 
As regards the set of the free parameters $\cal F$ only one case is considered when at a given $l$ all possible $m$ and $n$ values are taken into account  
i.e.
\begin{eqnarray}
&{\cal F}=\{[0,0,0]\}\cup \{[l,m,n]\vert l=2,\ldots,L,\ m=0,\ldots,l\ n=0,2,4\ldots,m\}\nonumber\\
&\setminus\{[l,0,0],[l,1,0]\vert  l=2,\ldots,L,\}.
\end{eqnarray}
In this case for a given odd $L$ the number of CF's is
$(21-14 L+15 L^2+2 L^3)/24$ and if $L$ is even then the number of CF's is $(24-14 L+15 L^2+2 L^3)/24$.

The Tables contain the ground state energy of the He atom  calculated by our models. 
The results of the simple models and the CF-Z-I and CF-Z-II descriptions are displayed in Table I. 
As it was expected the CF-S1 and CF-S2  descriptions give pure energies. Since the CF-S2 model contains functions of the form $w_{k,k,0}$ its results better than the model CF-S1. The  CF-Z-I and CF-Z-II descriptions contain all 
the possible CF's and the accuracy of the results are drastically improved. For large values of $L$ the descriptions CF-Z-I and CF-Z-II practically give the same results.

Further distinct improvement appears if the parameter of the exponential function of the trial function deviates from the value of $Z$. The results of the descriptions CF-$\alpha$-I and CF-$\alpha$-II are showed in Table II.  The energy gains are substantial if the value of the parameter $\alpha$ is not fixed 
to $Z$ but can be taken as a variational parameter. For these descriptions too it can be observed that for  large $L$ values the use the first- and second-type CF
give almost identical results. 
According to our knowledge of the literature the best energy for the ground state of the He atom is $-2.90360$ a.u. \cite{Rod07} with such a trial function where the cusp conditions are exactly fulfilled using trial functions of special forms. This trial function with energy  $-2.90360$ \cite{Rod07} contains 29 linear  variational parameters our CF-$\alpha$-II model gives energy -2.903706 and the number of linear parameters is only 14. 

\begin{center}
\begin{table}\label{cfenergy}
\caption{The same as Table I. The value of the parameter of the exponential function 
$\alpha$ is optimized.}
\begin{tabular}{|l|l||l|l|l|l|l|l||l|l|l|l|l|l|l|l|}
\hline
L&N&\multicolumn{6}{|c|}{CF-$\alpha$-I} &\multicolumn{8}{c||}{CF-$\alpha$-II}\\
\hline
&&$m_0$&$n_0$&$m_1$&$n_1$&\multicolumn{1}{|c|}{$\alpha$}&\multicolumn{1}{|c||}{E}&$m_0$&$n_0$&$m_\alpha$&$n_\alpha$&$m_\beta$&$n_\beta$&\multicolumn{1}{|c|}{$\alpha$}&\multicolumn{1}{|c|}{E}\\
\hline
1&1&2&0&2&2&1.956261&-2.877782&2&2&2&0&3&2&1.999999&-2.876582\\
2&3&2&0&2&2&1.999868&-2.898783&2&2&2&2&3&2&2.106178&-2.903345\\
3&7&2&0&3&0&2.073837&-2.903527&2&0&2&2&5&0&2.109632&-2.903526\\
4&14&2&0&3&0&2.093215&-2.903651&5&2&2&2&6&2&2.290487&-2.903706\\
5&24&2&0&3&0&2.241320&-2.903717&2&2&2&0&4&0&2.333256&-2.903717\\
6&38&2&0&7&6&2.345606&-2.903720&7&4&3&2&8&4&2.418335&-2.903722\\
7&56&2&0&3&0&2.413501&-2.903723&2&2&2&0&5&0&2.481257&-2.903723\\
8&79&6&2&6&4&2.555501&-2.903724&9&8&3&2&10&10&2.610711&-2.903724\\
\hline
\end{tabular}
\end{table}
\end{center}

In analogy with the double and triple basis set methods \cite{Dra02,Dra99} in order to improve the trial function the following form of  ansatz
\begin{equation}\label{dcf}
\sum_k \exp(-\alpha_k s)\sum_{[l,m,n]\in {\cal F}} C_{k,l,m,n}\tilde P_{l,m,n}(s,t,u)
\end{equation}
is introduced. It is obvious that the  wave function (\ref{dcf}) fulfills the cusp conditions. 
The free variables to be determined by diagonalization are $C_{k,l,m,n}$.
The use of trial function of the form (\ref{dcf}) may be called  double or triple cusp function description  when two or three different $\alpha$ parameters are used. In the numerical examples two $\alpha$ parameters are considered. Their values are optimized.
When two $\alpha$ parameters are used in the calculations and they are carried out with the first- and second-type  CF's 
they are called DCF-$\alpha$-I and DCF-$\alpha$-II descriptions, respectively. The results of these type of calculations are displayed in Table III. Only the $\alpha$ parameters are optimized, the auxiliary parameters are taken from Table II. Comparing the results of the three Tables it 
can be observed that the DCF-$\alpha$-I and DCF-$\alpha$-II descriptions give the best energy at a given number of basis size. Six decimal digits accuracy can be achieved with these models using only 48 basis functions.
\begin{center}
\begin{table}\label{cfenergy2}
\caption{Same as Table I. The same auxiliary parameters are used as in Table II but
double CF sets are used. The optimal values for $\alpha_1$ and $\alpha_2$ are shown.}
\begin{tabular}{|l|l||l|l|l|l|l|l|}
\hline
L&N&\multicolumn{3}{|c|}{DCF-$\alpha$-I} &\multicolumn{3}{c|}{DCF-$\alpha$-II}\\
\hline
&&\multicolumn{1}{|c|}{$\alpha_1$}&\multicolumn{1}{|c|}{$\alpha_2$}&\multicolumn{1}{|c|}{E}
&\multicolumn{1}{|c|}{$\alpha_1$}&\multicolumn{1}{|c|}{$\alpha_2$}&\multicolumn{1}{|c|}{E}\\
\hline
1&2&2.172266&2.493045&-2.892378&0.939756&2.001543&-2.876617\\
2&6&2.056165&2.210420&-2.903466&1.978561&2.897833&-2.903597\\
3&14&2.049804&2.729976&-2.903696&1.884354&2.804104&-2.903691\\
4&28&1.821203&3.241932&-2.903723&2.139435&2.670425&-2.903722\\
5&48&1.974419&3.762036&-2.903724&2.215298&3.218170&-2.903724\\
\hline
\end{tabular}
\end{table}
\end{center}

\section{Conclusions}\label{sect4}
New type of function sets are introduced. The cardinalities of the sets can be arbitrary large finite integer numbers. Superpositions of the new functions from the sets exactly fulfill the Kato's cusp conditions. The energy eigenvalues can be calculated with the standard matrix diagonalization technique and the more difficult minimalization 
with constraint approach to satisfy the cusp conditions can be avoided.  The new functions are special linear combinations of the basic Kinoshita-type terms. 
The set of the new functions can be changed versatilely.
Obviously the introduced method is valid starting from Hyllaraas-type basic terms. In this case numerical examples are given by calculating the ground state energy of the He atom. For a CF set approximately with fifty basis size the accuracy of the energy is six decimal digits and the cusp conditions are exactly fulfilled.

\section{acknowledgements}
This work was supported by the Hungarian Scientific
Research Fund NKFIH K112962 and I. H.
acknowledges partial support by Nazarbayev University (grant 090118FD5345).

\appendix*
\section{Derivation of the explicit solution for $k_{l,0,0}$}\label{proof}

The solution of the recursion (\ref{cuspsolz}) is searched in the form
\bea\label{recsol}
k_{l,0,0}=f_0(l)k_{0,0,0}+\overset {l} {\underset {i=2}  \sum} f(l,i) B_i  ~~~l\ge 2,
\eea
where $f_0(l)$ and $f(l,i)$ are unknown functions. 
Substituting (\ref{recsol}) into (\ref{cuspsolz}) using $k_{1,0,0}=\bar\alpha k_{0,0,0}$ and collecting the coefficients of $k_{0,0,0}$ and $B_i$ the following set of relations occur
\bea
&&f_0(l+2)-\frac{1}{l+2}\left(\bar\alpha-\frac{l+1}{2}\right)f_0(l+1) 
-\frac{\bar\alpha}{2(l+2)}f_0(l)=0  ~~~l\ge 2,\label{sol1} \\
&&f(l+2,i)-\frac{1}{l+2}\left(\bar\alpha-\frac{l+1}{2}\right)f(l+1,i) 
-\frac{\bar\alpha}{2(l+2)}f(l,i)=0 ~~~i\le l,l\ge 2,\label{sol2} \\
&&f(l+2,l+1)-\frac{1}{l+2}\left(\bar\alpha-\frac{l+1}{2}\right)f(l+1,l+1)=0  ~~~l\ge 1,\label{sol3} \\
&&f(l+2,l+2)=1  ~~~l\ge 0,\label{sol4} \\
&&f_0(2)=\frac{\bar\alpha^2}{2}, \label{sol5} \\
&&f_0(3)=\frac{\bar\alpha^3}{6}. \label{sol6}
\eea

From the explicit solution (\ref{expsolz}) it is easy to see that
\bea\label{f0sol}
f_0(l)=\frac{\bar\alpha^l}{l!}  ~~~l\ge 2,
\eea
and 
\bea\label{fsol}
f(l,i)=\frac{(-1)^{l+i}2^{i-l}i}{l!}
\left((l-1)!+\sum_{k=1}^{l-i}(-1)^k\bar\alpha^k(l-k-1)!2^k\right) ~~~l\ge 2, i \le l.
\eea
It remains to show that these functions satisfy the equations (\ref{sol1})-(\ref{sol6}).

Equations (\ref{sol5}) and (\ref{sol6}) are trivially fulfilled. 
From (\ref{f0sol}) we get $f_0(l+1)=\frac{\bar\alpha}{l+1}f_0(l)$ and
 $f_0(l+2)=\frac{\bar\alpha^2}{(l+1)(l+2)}f_0(l)$. Taking into account these formulas it is easy to check that (\ref{sol1}) is fulfilled.
 
Substituting (\ref{fsol}) into the l.h.s of (\ref{sol2}) we get 
\bea\label{tmpa1}
&&\frac{(-1)^{l+i}i2^{i-l}}{l!(l+2)}
\bigg[\frac{1}{4(l+1)} \bigg((l+1)!+\overset {l+2-i} {\underset {k=1} \sum} (-1)^k 
\bar\alpha^k (l-k+1)!2^k\bigg) \nonumber \\ 
&&+\left(\bar\alpha-\frac{l+1}{2}\right)\frac{1}{2(l+1)}
\bigg(l!+\overset {l+1-i} {\underset {k=1} \sum} (-1)^k 
\bar\alpha^k (l-k)!2^k \bigg)\nonumber \\
&&-\frac{\bar\alpha}{2}\bigg((l-1)!+\overset{l-i} {\underset {k=1} \sum} (-1)^k 
\bar\alpha^k (l-k-1)!2^{k}\bigg)\bigg].
\eea
The expression inside the square bracket in (\ref{tmpa1}) is a polinomial of $\bar\alpha$  with order $l+2-i$.
The constant term is
\bea
\frac{(l+1)!}{4(l+1)}-\frac{l+1}{2}\frac{l!}{2(l+1)}.
\eea
The coefficient of $\bar\alpha^n$ is
\bea
\frac{-l!2}{4(l+1)}+\frac{l!}{2(l+1)}-\frac{l+1}{2}\frac{1}{2(l+1)}(-1)(l-1)!2-\frac{(l-1)!}{2}
\eea
if $n=1$, 
\bea
&&\frac{(-1)^n}{4(l+1)}(l-n+1)!{2^{n}}+\frac{(-1)^{n-1}}{2(l+1)}(l-(n-1))!2^{n-1} \nonumber \\
&&-\frac{l+1}{2}\frac{1}{2(l+1)}(-1)^n(l-n)!2^n-\frac{1}{2}(-1)^{n-1}(l-(n-1)-1)!2^{n-1} 
\eea
if $1<n<l+2-i$
and  
\bea
\frac{(-1)^{l+2-i}}{4(l+1)}(l-(l+2-i)+1)!{2^{l+2-i}}+\frac{(-1)^{l+2-i-1}}{2(l+1)}(l-(l+2-i-1))!2^{l+2-i-1} 
\eea
if $n=l+2-i$. 
Since the last four expressions are identically equal with zero so the square bracket in (\ref{tmpa1}) is equal to zero too. All this means that (\ref{sol2}) is satisfied by the functions of (\ref{fsol}).

Using (\ref{fsol}) the first term of the l.h.s of (\ref{sol3}) is of the form
\bea
\frac{-(l+1)}{(l+2)!2} \bigg[(l+1)!-\bar\alpha l!2 \bigg]
\eea 
and the second term of the l.h.s. of (\ref{sol3}) looks like
\bea
-\frac{1}{l+2}\left(\bar\alpha-\frac{l+1}{2}\right)\frac{l+1}{(l+1)!}l!.
\eea
The comparision of  these terms show that (\ref{sol3}) is fulfilled.

Substituting (\ref{fsol}) into the l.h.s of (\ref{sol4}) it is trivial to see that  (\ref{sol4}) is fulfilled.
All required relations  (\ref{sol1})-(\ref{sol6}) are proved.

\end{document}